\documentclass[sigplan,nonacm]{acmart}
\makeatletter
\def\@ACM@checkaffil{
    \if@ACM@instpresent\else
    \ClassWarningNoLine{\@classname}{No institution present for an affiliation}%
    \fi
    \if@ACM@citypresent\else
    \ClassWarningNoLine{\@classname}{No city present for an affiliation}%
    \fi
    \if@ACM@countrypresent\else
        \ClassWarningNoLine{\@classname}{No country present for an affiliation}%
    \fi
}
\makeatother
\renewcommand\footnotetextcopyrightpermission[1]{}

\usepackage{xspace}
\usepackage{graphicx}
\usepackage{subcaption}
\usepackage{pgfplots}
\usetikzlibrary{patterns}
\pgfplotsset{compat=1.18}

\usepackage[small,compact]{titlesec}
\usepackage[font={small}]{caption}  
\usepackage{balance} 

\acmPrice{}
\acmDOI{}
\acmYear{2024}
\acmISBN{}
\acmConference{}{}{}
\acmBooktitle{}
\begin{document}

\newcommand{\rust}[1]{\ensuremath{\mathtt{#1}}\xspace}
\newtheorem{claim}{Claim}
\newtheorem{observation}{Observation}
\newtheorem{property}{Property}
\newtheorem{invariant}{Invariant}

\newcommand{\namedref}[2]{\hyperref[#2]{#1~\ref*{#2}}}
\newcommand{\theoremref}[1]{\namedref{Theorem}{#1}}
\newcommand{\equationref}[1]{\hyperref[#1]{(\ref*{#1})}}
\newcommand{\invariantref}[1]{\namedref{Invariant}{#1}}
\newcommand{\lemmaref}[1]{\namedref{Lemma}{#1}}
\newcommand{\claimref}[1]{\namedref{Claim}{#1}} 
\newcommand{\obsref}[1]{\namedref{Observation}{#1}} 
\newcommand{\observationref}[1]{\namedref{Observation}{#1}}
\newcommand{\corollaryref}[1]{\namedref{Corollary}{#1}} 
\newcommand{\propertyref}[1]{\namedref{Property}{#1}} 
\newcommand{\figureref}[1]{\namedref{Figure}{#1}}
\newcommand{\algorithmref}[1]{\namedref{Algorithm}{#1}}
\newcommand{\appendixref}[1]{\namedref{Appendix}{#1}}
\newcommand{\tableref}[1]{\namedref{Table}{#1}}
\newcommand{\sectionref}[1]{\namedref{Section}{#1}} 
\newcommand{\appref}[1]{\namedref{Appendix}{#1}} 
\newcommand{\definitionref}[1]{\namedref{Definition}{#1}} 
\let\lineref\linerefa
\newcommand{\lineref}[1]{\namedref{Line}{#1}} 
\newcommand{\algoref}[1]{\namedref{Algorithm}{#1}} 

\newcommand{\maybeshow}[1]{#1}

\newcommand{\Rati}[1]{\textcolor{red}{\textbf{Rati:} \maybeshow{#1}}}
\newcommand{\Sasha}[1]{\textcolor{orange}{\textbf{Sasha:} \maybeshow{#1}}}
\newcommand{\George}[1]{\textcolor{blue}{\textbf{George:} \maybeshow{#1}}}
\newcommand{\Daniel}[1]{\textcolor{purple}{\textbf{Daniel:} \maybeshow{#1}}}
\newcommand{\Zi}[1]{\textcolor{pink}{\textbf{Zi:} \maybeshow{#1}}}
\newcommand{\Igor}[1]{\textcolor{olive}{\textbf{Igor:} \maybeshow{#1}}}
\newcommand{\Nuno}[1]{\textcolor{teal}{\textbf{Nuno:} \maybeshow{#1}}}
\newcommand{\remove}[1]{}

\newcommand{\DeltaLane}{RapidLane}
\newcommand{\Aptos}{Aptos}

\newcommand{\vwave}{\rust{validation\_wave}}
\newcommand{\ttwave}{\rust{triggered\_wave}}
\newcommand{\trwave}{\rust{required\_wave}}
\newcommand{\tvwave}{\rust{validated\_wave}}
\newcommand{\cwave}{\rust{commit\_wave}}
\newcommand{\cidx}{\rust{commit\_idx}}

\title{Deferred Objects to Enhance Smart Contract Programming with Optimistic Parallel Execution}

\author{George Mitenkov}
\affiliation{\institution{Aptos Labs}}
\author{Igor Kabiljo}
\affiliation{\institution{Aptos Labs}}
\author{Zekun Li}
\affiliation{\institution{Aptos Labs}}
\author{Alexander Spiegelman}
\affiliation{\institution{Aptos Labs}}
\author{Satyanarayana Vusirikala}
\affiliation{\institution{Aptos Labs}}
\author{Zhuolun Xiang}
\affiliation{\institution{Aptos Labs}}
\author{Aleksandar Zlateski }
\affiliation{\institution{Aptos Labs}}
\author{Nuno P. Lopes}
\affiliation{\institution{INESC-ID / IST - University of Lisbon}}
\author{Rati Gelashvili}
\affiliation{\institution{Aptos Labs}}
\renewcommand{\shortauthors}{Mitenkov, Kabiljo, Li, Spiegelman, Vusirikala, Xiang, Zlateski, Lopes, and Gelashvili}

\begin{abstract}




One of the main bottlenecks of blockchains is smart contract execution.
To increase throughput, modern blockchains try to execute transactions in parallel.
Unfortunately, however, common blockchain use cases introduce read-write conflicts between transactions, forcing sequentiality.




We propose \DeltaLane{}, an extension for parallel execution engines that allows the engine to capture computations in conflicting parts of transactions and defer their execution until a later time, sometimes optimistically predicting execution results.
This technique, coupled with support for a new construct for smart contract languages, allows one to turn certain sequential workloads into parallelizable ones.

We integrated \DeltaLane{} into Block-STM, a state-of-the-art parallel execution engine used by several blockchains in production, and deployed it on the \Aptos{} blockchain.
Our evaluation shows that on commonly contended workloads, such as peer-to-peer transfers with a single fee payer and NFT minting, \DeltaLane{} yields up to $12\times$ more throughput.

\end{abstract}

\maketitle

\section{Introduction}
\label{sec:introduction}

A blockchain is a distributed system that consists of two sets of nodes: validators and clients.
Clients submit transactions to validators.
Validators, in turn, agree on a sequence of submitted transactions, and execute them in a deterministic way, thereby changing the state of the blockchain.

While traditionally being designed as distributed ledgers supporting only simple transfers of digital currencies, modern blockchains have a significantly broader scope of applications thanks to smart contracts --- user-defined programs stored on the blockchain as bytecode~\cite{kemmoe2020, zibin2020}.
Popular smart contracts include automated market makers~\cite{mohan2022}, financial derivatives~\cite{paulsonluna2020, zhang2021}, minting non-fungible tokens (NFTs)~\cite{arora2022,patel2024}, automatic insurance settlement~\cite{jouini2023}, and oracles to provide external off-chain data~\cite{beniiche2020}.
 
To unlock many other use cases, blockchains need to overcome several challenges,
namely low throughput. 
Prominent blockchains such as Bitcoin~\cite{nakamoto2009} and Ethereum~\cite{wood2014} 
  can only process $7$ and $30$ transactions per second (TPS), respectively, compared to more than 65k TPS processed by Visa~\cite{visa2017}.

A crucial component to increase throughput, which many modern blockchains have introduced, is a parallel execution engine.
This allows a single validator node to process transactions batched in a block in parallel~\cite{fouda2024}.
Unfortunately, some of the core blockchain workloads are inherently highly contended~\cite{heimbach2024}, creating read-write conflicts between different transactions in a block (i.e., conflicts where one transaction writes to and another transaction reads form the same location in the state).

For example, many blockchains have the concept of sponsored transactions~\cite{aptosdev2024, eip3074, suidev2024}.
Instead of having the users pay the transaction fees, the fees are payed by sponsors (also called fee payers).
This mechanism is used by applications that want to cover the costs for their users, e.g., for registration.
For a transaction to execute successfully, the sponsor must have sufficient funds to cover the fees, which will be charged (subtracted from the sponsor's account balance).\footnote{Transaction fees are used to protect blockchains from denial-of-service (DoS) attacks and to cover
the infrastructure costs.
They are based on the amount of resources used
and are computed during the transaction execution 
to protect from infinite loops.
}
Because many such transactions can be issued within a short time to interact with the same application, they all exhibit read-write conflicts, creating contention.
Other examples include direct transfers to or from a single account (crediting and debiting the same balance), total token supply tracking (every transaction updates the supply when burning a fraction of the fee~\cite{ethereum2024}), and NFT minting (keeping a counter to limit the maximum number of NFTs that can be minted).

In order to avoid the problem of read-write conflicts, some blockchains alter their systems, either removing contention altogether (e.g., forgoing the ability to track the total supply at any given moment in their smart contracts~\cite{solana2024}), or scheduling the conflicting transactions across multiple blocks~\cite{solana2017}.
While such scheduling can lead to higher parallelism within a single block, it has a substantial impact on user experience, as some users may need to wait a significant amount of time until their transaction is scheduled for execution.
Moreover, it is in general not possible to accurately predict what data a transaction will access (except for blockchains that require contract developers to manually specify an over-approximation of the data that will be read and written).

At the same time, users currently do not have to pay higher fees if they execute a contract that causes high contention.
As a result, smart contract developers have no incentive to avoid data conflicts (except potentially for blockchains that charge a fee based on the specified read/write sets).

Together with the lack of guidelines and the absence of good developer documentation, the result is poorly-written, sequential smart contracts, which cannot be efficiently executed by validators even when using sophisticated parallel execution engines.
Furthermore, existing smart contract programming languages do not have the notion of parallelism, e.g., they do not support atomic types.
Even if developers do want to avoid contention, they have to come up with workarounds to ensure that it is possible to execute multiple transactions at the same time.

There exist multiple well-known techniques for optimizing programs for contended workloads, including but not limited to sharded counters~\cite{herlihy1995} or exploiting commutativity of operations~\cite{kulkarni2011}.
However, these techniques do not suit well the typical blockchain workloads, as they often require giving up key properties of computations.
Reading contented data is often unavoidable.

For example, the balance of a fee payer's account must be read to know it has sufficient funds, as otherwise it can make the network vulnerable to DoS attacks.  
Another example is NFT minting.
A collection of NFTs is usually minted with a specified limit on the number of tokens that can ever exist and the index of the minted token can be used as part of its name, e.g., ``arXiv \#2024'' (a collection of `arXiv' tokens, with the particular token having index 2024).

\paragraph{Contributions}

The main observation of this paper is that a large fraction of the read-write conflicts from smart contracts do not have to be qualified as such.
Usually, conflicting transactions either update a value but do not need the result, or read the value before the update to check if some precondition is satisfied  (e.g., if the balance is large enough).

As a result, parallel execution engines can defer computations on contended data for every transaction until a later time, without being required to create a read-write conflict by reading the intermediate values.
In case a precondition must be evaluated, we find that it is usually sufficient to predict the result and validate it at a later time.

\begin{figure}[t]
    \includegraphics[width=0.95\linewidth]{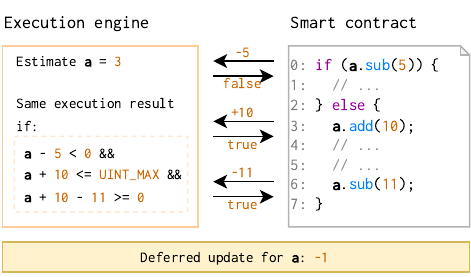}
    \caption{On the right, a snippet of smart contract code which is being executed.
    The code uses a deferred object \rust{a} which supports checked arithmetic via the \rust{add} and \rust{sub} methods.
    Both methods return \rust{true} if there is no overflow or underflow.
    On the left, a parallel execution engine which captures the updates to \rust{a} from the contract ($-5$, $+10$, and $-11$).
    Instead of reading the value of \rust{a}, its value is estimated to make a prediction about the outcomes of arithmetic operations.
    Also, modifications to the value stored inside the deferred object are delayed, thus eliminating read-write conflicts if the object is shared across multiple transactions.
    }
    \label{fig:introduction-solution}
\end{figure}

The main contributions of this paper are:

\begin{itemize}
    \item Deferred objects --- a new concept for smart contract programming languages that enables developers to defer computations.
    With deferred objects, developers can explicitly specify objects which are expected to be frequently mutated by transactions in parallel.
    Importantly, multiple transactions updating the same deferred object do not conflict, and the updates do not have to be commutative.
    For example, Figure~\ref{fig:introduction-solution} shows a snippet of a smart contract that uses deferred objects to avoid read-write conflicts on a shared counter, and how it can be used by the execution engine.
    \item \DeltaLane{} --- a production-grade parallel execution engine extension used to implement deferred objects.
    \DeltaLane{} is built on top of Block-STM~\cite{gelashvili2023}, a state-of-the-art shared-memory parallel execution engine adopted by multiple blockchains~\cite{aptos2022, monad2024, polygon2024, sei2024, starknet2024}.
    Thanks to deferred semantics, conflicting parts of transactions are captured and deferred in dedicated lanes, where they are also performed efficiently in parallel.
    To increase parallelism, the execution results are optimistically predicted as depicted in Figure~\ref{fig:introduction-solution}.

    \item A new open-source benchmark suite containing common blockchain transactions, including real contracts for peer-to-peer transfers, sponsored transactions, and NFT minting.
    The benchmark suite also includes synthetic benchmarks with conflicting workloads designed to stress test parallel execution engines.
    This benchmark suite allows drawing more qualitative comparisons across blockchains over single TPS numbers.
    
    \item A thorough evaluation of \DeltaLane{} on the \Aptos{} blockchain.
    We study how our system performs on the benchmark suite, and additionally, conduct a series of smaller experiments to describe the system's behavior.
\end{itemize}

As a secondary contribution, the paper enhances Block-STM with a rolling commit mechanism.
At virtually no performance cost, it allows the system to know when each transaction is committed, as opposed to block granularity in Block-STM's original lazy commit mechanism.
\section{Background}\label{sec:background}

\subsection{Blockchains and smart contracts}

There are two types of nodes in a blockchain: validators and clients.
Validators maintain the global state of the blockchain, as well as the transaction history that records all transactions executed so far, and the corresponding execution outcomes.

Clients interact with the blockchain by submitting transactions to validators.
Validators agree on an sequence of blocks, each containing an ordered list of transactions submitted by clients.
Then, validators execute the transactions of each block locally according to the specified order.
After execution, all validators must reach the same global state.

Conceptually, the global state is a key-value store, where the key is known as an address, and the value can be either data or smart contract code. %
Data entries can be arbitrarily complex data structures, and are stored in a serialized format.

\begin{figure*}[t]
    \includegraphics[width=\textwidth]{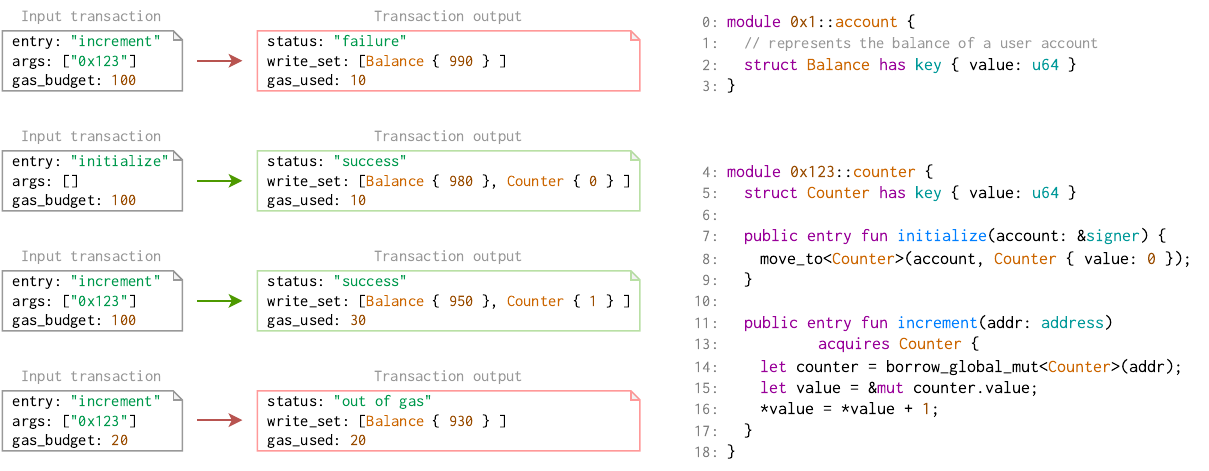}
    \caption{
        On the right, an example of a smart contract in the Move programming language which implements a simple counter (lines 4--18).
        Functions \rust{initialize} and \rust{increment} are used to, respectively, create a new counter and modify it.
        On the left, an example of four user transactions and their execution output, from top to bottom:
        1) Transaction fails to increment the counter because it does not exist in the global state.
        Gas is still charged, updating the user's balance in the write-set.
        2) Transaction initializes a counter.
        3) Transaction successfully increments the counter. 
        4) Transaction runs out of gas due to insufficient gas budget. The counter is not modified, but gas is still charged.
    }
    \label{fig:background-vm-smart-contract-example}
\end{figure*}

Smart contracts are user-defined programs, typically implemented in domain-specific languages such as Solidity~\cite{solidity2024}, Vyper~\cite{vyper2020}, or Move~\cite{blackshear2019} (Figure~\ref{fig:background-vm-smart-contract-example}).
Some blockchains also allow subsets of general-purpose programming languages such as Haskell, JavaScript, or Rust~\cite{cardano2024, near2021, solana2017}.

Smart contracts are stored as (serialized) bytecode, which ensures portability and allows contracts to be executed on any platform irrespective of the hardware vendor.
The bytecode used depends on the blockchain and/or the smart contract language.
A contract can be deployed to the blockchain by submitting a transaction with the contract's bytecode.

\subsection{Virtual machines and smart contract execution}

Clients submit transactions specifying which smart contract to execute in an isolated virtual machine (VM).
The isolation is crucial, as it means that neither any temporary state nor the execution itself can be affected from outside of the VM.

During execution, transactions can read and modify data from the global state.
Instead of making changes to the state directly, modifications made by each transaction are recorded as a write-set, i.e., a set of all stores. 

Since smart contracts can take arbitrary long to execute (e.g., they can contain an infinite loop), VMs limit the amount of resources transactions can use.
Usually this is done by associating each bytecode instruction with a cost called gas.

When clients submit transactions, they specify a gas budget, i.e., the maximum amount of gas they are willing to pay to execute the transaction.
VMs keep track of gas usage when running smart contracts and, as a result, every transaction eventually terminates (possibly by depleting the gas budget).

The output of a transaction that executes successfully and within the gas budget is a write-set and the gas used.
A transaction fails if it exceeds the gas budget or a runtime error occurs, such as division by zero.
Because each transaction is an atomic unit, changes made by failed transactions' code are discarded and are not included in the write-set.

\subsection{Parallel execution engines}

Transactions are executed by validators in blocks using so-called execution engines.
An execution engine takes an ordered block of transactions, and returns the outputs for each transaction. 
Then, validators produce the new global state by applying the write-sets to the pre-block state.
Also, all transaction outputs are added to the transaction history ledger.

Modern execution engines process transactions in a block in parallel.
Importantly, all transaction outputs must be deterministic and always equal to the outputs of transactions as if they were executed sequentially in the given order.
Parallel execution engines can be classified into two groups based on the different kind of parallelism they use: static and dynamic.

Execution engines that use static parallelism, e.g., used by Solana~\cite{solana2017} and Sui~\cite{sui2022}, require users to explicitly annotate their transactions with the values a transaction may read from or write to.
Then, the transactions which do not have declared data access conflicts are executed in parallel, while the conflicting ones are executed sequentially.

In contrast, some blockchains such as Aptos~\cite{aptos2022} and Polygon~\cite{polygon2024} use dynamic
parallelism.
Their execution engines optimistically execute all transactions in parallel, hoping that there are no conflicts.
When a conflict is detected, the conflicting transactions are re-executed.
On workloads with many conflicts, both static and dynamic approaches perform poorly: the static yields a sequential execution, while the dynamic ends up re-executing transactions many times.

Block-STM~\cite{gelashvili2023} is a state-of-the-art parallel execution engine for smart contracts.
It implements dynamic parallelism using multi-versioning and optimistic concurrency control.
Transactions are optimistically processed in parallel (execution), and then checked for possible conflicts (validation).
Block-STM maintains a read-set (all values read from the global state) to detect changes in the read values.
If some value changed, a read-write conflict is discovered and the transaction is scheduled for re-execution.

Block-STM uses multi-version data-structures to avoid write-write conflicts~\cite{bernstein1983}.
Each transaction execution is associated with a version and the data structure contains a slot per version for every modified key in the global state.
To read from the data structure, a transaction with version $v$ takes the value from the slot associated with the highest version that is smaller than $v$, thus eliminating write-write conflicts.
Also, since transactions run in a sandboxed VM, Block-STM does not require opacity~\cite{guerraoui2008} or privatization~\cite{spear2007}. 

\section{Deferred objects: a parallel programming model for smart contracts}
\label{sec:aggregators}

\begin{figure}[t]
    \includegraphics[width=\linewidth]{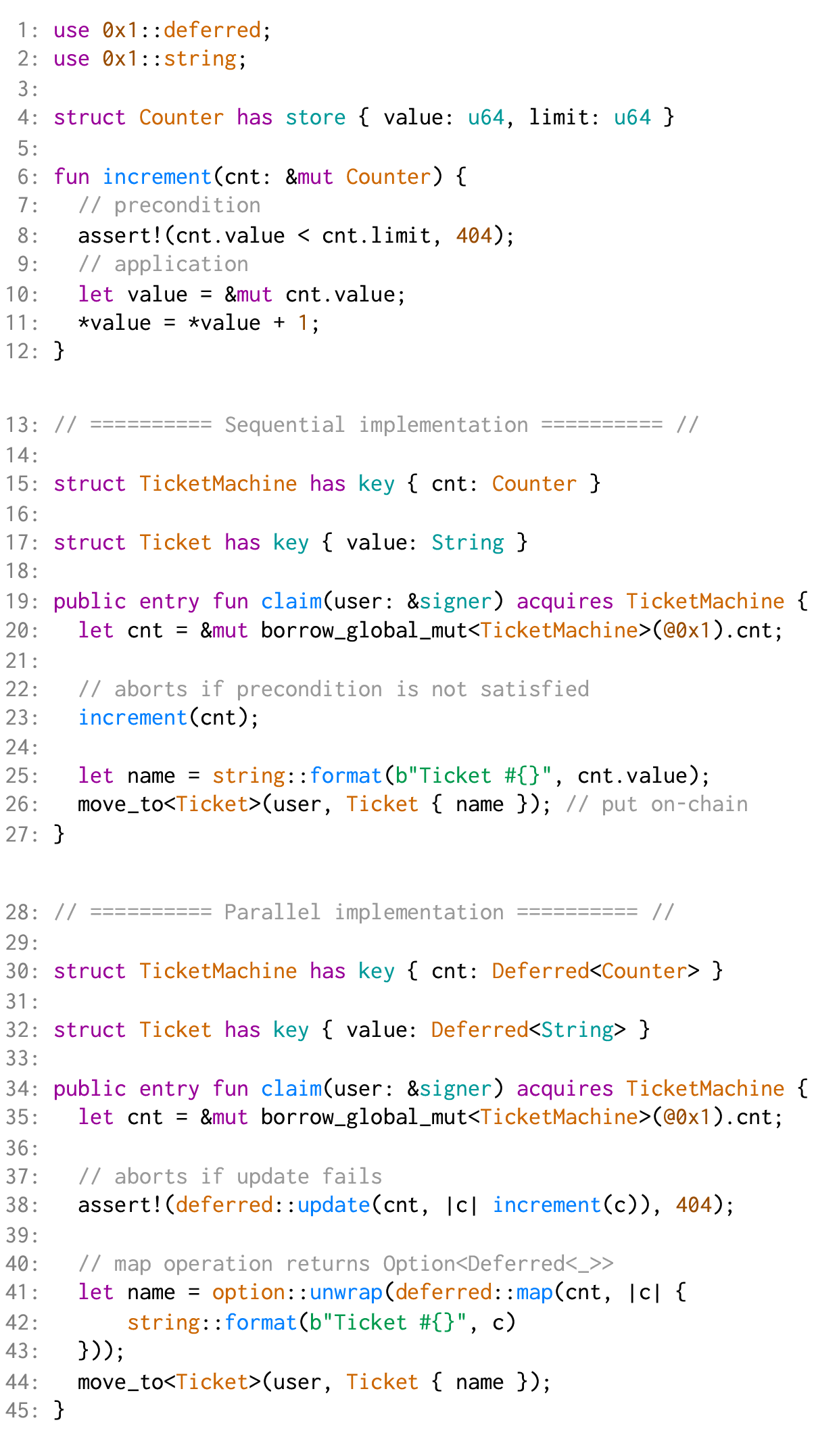}
    \caption{Example of a smart contract written in Move that implements logic to sell tickets.
    Users call the \rust{claim} function, and if tickets are still available, a \rust{Ticket} is stored in the global state under their address.
    If all tickets are sold out, the call aborts.
    The figure shows two different implementations of this smart contract: without (lines 13--27) and with (lines 28--45) deferred objects.}
    \label{fig:aggregators-example}
\end{figure}

We propose an extension to programming languages and VM runtimes for smart contracts: \emph{deferred objects}, which are containers that capture and defer computation.
In this section, we present the concept of deferred objects with an example.
We consider the Move smart contract programming language, which encodes logic to sell a limited number of concert tickets (see Figure~\ref{fig:aggregators-example}).
We use it to show how a sequential implementation with non-trivial read-write conflicts and non-commutative operations can be turned into a parallel one, thanks to a deferred counter.
As we go through the example, we also discuss the necessary interaction between the VM and the parallel execution engine.
In later sections, we define deferred objects formally and describe the required changes in the VM implementation.

\subsection{Deferred objects by example}

We first consider a sequential implementation of the contract (lines 13--27).
The entity that issues the tickets is represented by the \rust{TicketMachine} struct, which stores a counter to track the number of tickets issued so far.
A concert ticket is represented as a \rust{Ticket} struct.

Now, we assume there is a \rust{TicketMachine} stored in the global state, and that there are multiple transactions that call the \rust{claim} function in order to obtain a ticket for different users.
Clearly, transactions calling the \rust{claim} function cannot be executed in parallel:
each of them reads and writes the \rust{TicketMachine}'s global counter.

Next, we show how to use deferred objects to remove the read-write conflict when mutating the \rust{TicketMachine} in order to allow transactions calling the \rust{claim} function to be executed in parallel.
First, both \rust{Ticket} and \rust{TicketMachine} have their fields wrapped as deferred objects.
The deferred type specifies that the computations performed on these fields are captured and deferred until commit time.
A transaction that accesses a \rust{TicketMachine} (line 35), gets the counter wrapped into a deferred object, without leaking the actual value it stores.


Calling \rust{update} on the deferred object tries to apply the \rust{increment} function on the wrapped counter (line 38).
This function returns a boolean indicating whether the counter has been incremented.
In practice, the execution engine running this transaction detects that an update to a deferred object is being made, captures the computation $\lambda x\ . \ x + 1$, and decides if the \rust{increment} call succeeds or not (i.e., the counter's limit is exceeded).
The decision is made using a heuristic that does not require reading the most recent value of the counter.
Assuming the increment succeeds, \rust{update} returns \rust{true} and the engine logs the deferred computation.

Next, the name of the ticket needs to be computed without reading the counter value (lines 40-44).
The \rust{map} operation creates a new deferred string that produces the name of the ticket.
The call to \rust{format} is deferred until commit time, and a dummy string is returned to initialize the ticket (line 42).%
\footnote{
Note that deferring computations slows down the commit phase.
In this example, the overhead of delayed string creation can be non-negligible.
A more efficient alternative would be to have the ticket store just its number, and have the name computed on the fly in the ``getter'' function.
We use a deferred string here to show the generality of deferred objects.}
Finally, the \rust{Ticket} is published to the global state, and the transaction execution terminates.
The VM, in addition to the write-set with a single \rust{Ticket}, produces two deferred changes: $\lambda x\ . \ x + 1$ for the \rust{cnt} field, and string formatting for the \rust{name} field.
By deferring the read of the counter, and the subsequent computations, the write to the \rust{TicketMachine} has been eliminated.
As a result, transactions have no read-write conflicts and can execute in parallel.

However, the fields of \rust{Ticket} and \rust{TicketMachine} are not yet known when transactions finish and are about to be committed.
During commit, the execution engine calculates the correct value of the counter (by reading the initial value and applying the addition) and validates that the decision for the outcome of \rust{increment} was indeed correct.
Similarly, the name of the \rust{Ticket} is updated by applying the string formatting to the updated counter value.

\subsection{Deferred object operations}

We denote the type of a deferred object that stores an object of type $V$ as $D_V$.
Deferred objects support five operations:
\begin{itemize}
    \item $\mathit{create} : V \rightarrow D_V$. Constructs a deferred object from an input object, capturing it by value.
    \item $\mathit{reveal} : D_V \rightarrow V$. Returns the copy of the value stored in  the deferred object.
    \item $\mathit{update} : D_V \times (V \rightarrow V) \rightarrow \{\mathit{true}, \mathit{false}\}$. Updates the stored object using the input function, $f$, and returns \emph{true} if the precondition of $f$ is satisfied. Otherwise, returns \emph{false} without applying $f$.
    \item $\mathit{map} : D_U \times (U \rightarrow V) \rightarrow  \{ D_V, \varnothing \}$. Maps the value from the deferred object into a new object using the input function, $g$, and returns it wrapped as a deferred object if the precondition of $g$ holds. Otherwise, returns $\varnothing$.
    \item $\mathit{combine} : D_U \times D_V \times (U \times V \rightarrow W) \rightarrow  \{ D_W, \varnothing \}$. Combines the objects (possibly of different types) stored in a pair of deferred objects into a new object of type $W$ using the input function, $h$, and returns it wrapped as a deferred object if the precondition of $h$ is satisfied. Otherwise, returns $\varnothing$.
\end{itemize}

When going through the concert ticket example, we illustrated how \emph{update} and \emph{map} work and motivated their usage.
We now discuss the remaining functions.

The \emph{create} operation is crucial for correctness because it ensures that the stored object is captured by value, and so there are no mutable references which can mutate the stored object by means other than through the deferred object operations.
Using the concert ticket example, it is not possible to obtain \rust{\&mut} \rust{Counter} in any way.

Combining deferred objects is similar to a \emph{map}, but allows folding multiple deferred objects into a single one.
For example, this can be used to compute the average of a few counters in parallel without introducing read-write conflicts.

Lastly, \emph{reveal} returns a copy of the value stored in the deferred object and is used when a transaction needs to know the actual value straight away.
It creates read-write conflicts and therefore it is best avoided for performance.
We discuss the implications for execution engines in Section~\ref{sec:deltalane}.


\subsection{Deferred logs}

\begin{figure}[t]
    \includegraphics[width=\linewidth]{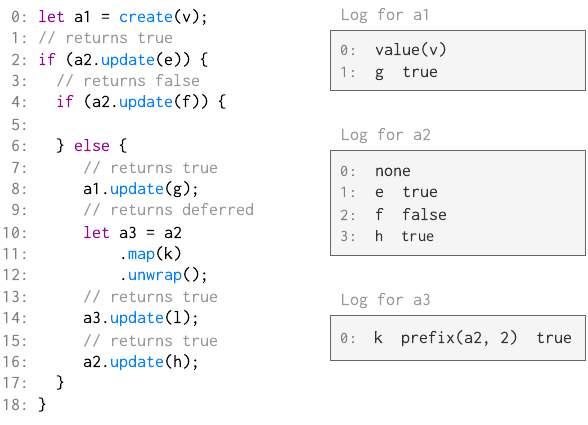}
    \caption{On the left, an example code snippet using three deferred objects \rust{a1, a2, a3} with comments that specify the estimated outcomes of preconditions during execution.
    On the right, the corresponding set of logs.
    Object \rust{a1} is created, and so row 0 of the log shows the initial value.
    Object \rust{a2} already existed (e.g., it was created by another transaction), and so row 0 of the log is \emph{none}.
    Because the precondition of \emph{f} failed, the corresponding log entry is \emph{false}.
    Object \rust{a3} is derived from \rust{a2} and is initialized with the prefix of the log of \rust{a2}.
    }
    \label{fig:aggregators-logs}
\end{figure}

To support deferred objects, the VM extends the transaction output with a set of \emph{logs} for every deferred object accessed during execution.
A log is a sequence of deferred operations.
Figure~\ref{fig:aggregators-logs} shows an example that uses three deferred objects, as well as the corresponding logs when executed.

The first row of the log corresponds to the deferred object's initial value.
If the deferred object is created with value $v$, \emph{value(v)} is logged.
If the deferred object is created with \emph{map}, `\emph{f prefix(a, i) c}' is logged, where $f$ is the mapping function, $c$ is the estimated outcome of its precondition, $a$ is the deferred object that was mapped, and $i$ is the size of the log of $a$ when \emph{map} was called.
Similarly, if the deferred object is created with \emph{combine} on deferred objects $a, b$, `\emph{f prefix(a, i) prefix(b, j) c}' is logged. 
If \emph{reveal} or \emph{update} are called, \emph{none} is logged.
The remaining rows are either a revealed value (logged as `\emph{revealed(v)}'), or tuples of functions applied with \emph{update}s and the outcomes of the estimated preconditions, logged during execution.

\subsection{Revealing deferred objects}

Sometimes the value stored in a deferred object is needed, either because the program calls \emph{reveal} explicitly or because the engine does not want to speculate on its value.
We now show how the VM computes this value.

\begin{figure*}[t]
    \includegraphics[width=0.8\textwidth]{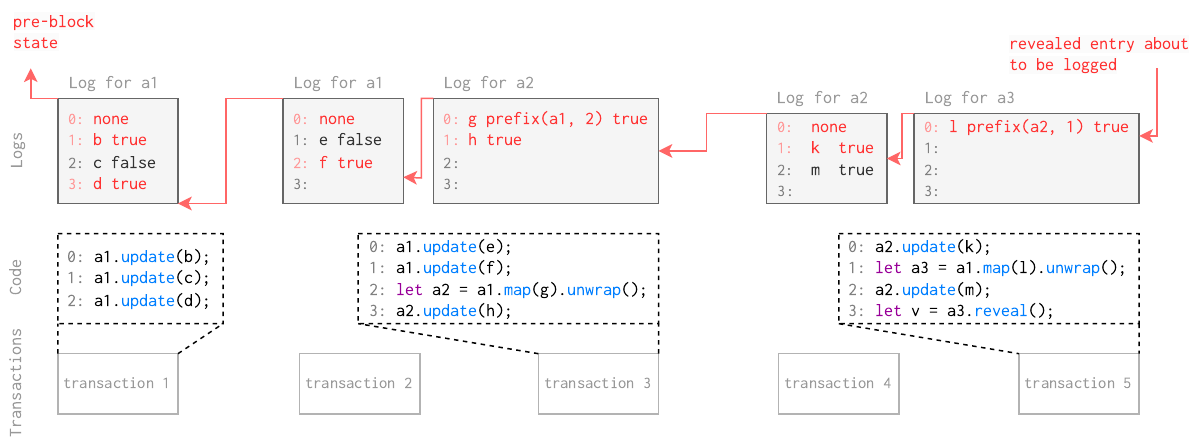}
    \caption{Example of revealing the deferred object \rust{a3} in transaction 5.
    Revealing traverses preceding transactions for deferred objects (in red) to identify all logs needed for replay.
    The initial value of the deferred object is retrieved from storage. 
    }
    \label{fig:aggregators-materialization}
\end{figure*}

First, we define a helper function \emph{replay(a, i, j)} which reveals the value of a deferred object $a$ used in transaction $\mathit{tx}_i$ after the $j$-th operation in the log has been performed.
Then, the value can be computed as follows:
\begin{enumerate}
    \item Compute the value stored in the deferred object before the current transaction $\mathit{tx}_i$.
    First, we check the initialization ($0$-th) row in $a$'s log.
    If it is \emph{value(v)}, do nothing.
    If it contains \emph{prefix(c, l)}, find $\mathit{tx}_k$ with the largest $k \le i$ such that it has a log for $c$.
    Then, replace the \emph{prefix} with \emph{value(replay(c, k, l))}.
    Finally, if it contains \emph{none}, find $\mathit{tx}_k$ with the largest $k < i$ such that it has a log for $c$, and replace it with \emph{value(replay(c, k, L))}, where $L$ is the size of the log.
    \item Replay the log for the current transaction applying only the functions which satisfy the precondition (have \emph{true} in the corresponding log entry).
    If `\emph{revealed(v)}' is encountered during replay, $v$ is returned immediately.
\end{enumerate}
Figure~\ref{fig:aggregators-materialization} gives an example of a reveal operation.

\subsection{Using deferred objects}

The purpose of deferred objects is to provide a flexible parallel programming model that developers can use to improve efficiency.
More specifically, for objects which can experience high contention, developers can:
\begin{enumerate}
    \item Wrap them as deferred objects, and use \emph{update} operations to reduce the number of read-write conflicts.
    \item Map, and combine deferred objects to avoid revealing where the shared object needs to be transformed without introducing read-write conflicts.
\end{enumerate}
The execution engine then only ensures that estimations are accurate and mispredicted transactions (and those that depend on them) are re-executed.

\section{\DeltaLane{}}\label{sec:deltalane}

In this section, we present \DeltaLane{}, an extension to the Block-STM parallel execution engine to optimize contended workloads.
\DeltaLane{} leverages deferred objects by capturing the logs provided by the VM for each transaction, and replaying them at commit time.
Thanks to the multi-version data structures of Block-STM, transactions that create logs for the same deferred objects do not conflict (similarly to how write-write conflicts are avoided).

In our implementation, we restrict the types that can be wrapped as deferred objects to integer counters and small-sized strings (up to 256 bytes).
The counter's value can be updated with addition or subtraction, as long as it is within some fixed lower and upper bounds.
We also allow for integer deferred objects to be mapped once.

\subsection{Block-STM preliminaries}
\label{sec:system:bstmprelim}

Block-STM takes a block of $n$ transactions $[\mathit{tx}_1, \mathit{tx}_2, \ldots, \mathit{tx}_n]$  as input, and returns $n$ transaction outputs.
The outputs are the same as if transactions were executed in the given order $\mathit{tx}_1 < \mathit{tx}_2 < \ldots < \mathit{tx}_n$, i.e., sequentially.

The write-set produced when a transaction is executed is recorded in a multi-versioned data-structure, \emph{MVMemory}, where writes from different transactions are stored independently. 
MVMemory also exposes an interface that lets $\mathit{tx}_i$ read the currently ``latest'' value for a given key, i.e., stored write from $\mathit{tx}_j$ with highest $j < i$ (or, in case there are no previous writes, from \emph{storage}, i.e., based on the pre-block global state).

In Block-STM, transactions are executed concurrently and thus the write-sets can be applied to MVMemory out of order. 
This makes reads \emph{speculative}, i.e., the results depend on the contents of MVMemory at the time of reading and may not yield the same result as if transactions were executed sequentially.
A component called \emph{CapturedReads} stores information about all reads done by transactions.
This information is later used to validate the correctness of the execution.

The \emph{scheduler} provides tasks to each worker thread.
An \emph{execution task} executes a transaction, with an \emph{incarnation number} equal to the number of times it has been executed previously.
It is guaranteed that incarnations of the same transaction occur in order and are never concurrent.

A \emph{validation task} validates the read-set of the transaction's latest incarnation, checking for conflicting writes done by transactions that are ordered before in the block, but which may have completed executing after the current one.
Since earlier transaction outputs may affect later transactions, the scheduler is designed to prioritize tasks with lower transaction indices.
Scheduling satisfies the following:

\begin{invariant} 
\label{inv:bstmvalidation}
Each incarnation is validated at least once.
After completion, an incarnation of a transaction $\mathit{tx}_i$ gets scheduled for (re-)validation whenever a task for transaction $\mathit{tx}_j$ for $j < i$ may influence the result of a read in the read-set of $\mathit{tx}_i$. 
\end{invariant}

When validation fails, the incarnation is \emph{aborted} and the transaction is scheduled for re-execution (with an incremented incarnation number).
Transactions with a higher number are also re-validated.
As a result, and as proved in~\cite{gelashvili2023}, Block-STM satisfies the following property:

\begin{property}
\label{prop:bstmvalidation}
For any $1 \leq i \le n$, the last incarnation of $\mathit{tx}_i$ is validated in a global state that is consistent with the sequential execution of transactions $1, 2, \ldots, i$ and never changes.  
\end{property}


In order to describe our changes to Block-STM in a way that is oblivious to the complex internal details, we define the following abstract API for the scheduler:
\begin{itemize}
    \item \rust{schedule\_execution(i)}, \rust{schedule\_validation(i)} to schedule $\mathit{tx}_i$ 
       for execution or validation.
    \item \rust{schedule\_suffix\_validation(i)} to schedule transactions $i, \ldots, n$ for validation.
\end{itemize}

These methods guarantee that after they are called, either transaction $i$ gets validated,
  transaction $i$ gets executed, or transactions $i,\ldots,n$ all get validated, respectively.
While \rust{schedule\_execution} and \rust{schedule\_validation} are more common 
  functionalities, \rust{schedule\_suffix\_validation} is specific to Block-STM (since the outputs of earlier transactions affecting the later ones).


Another important aspect of Block-STM is dependency management.
When an incarnation is aborted, its writes stored in MVMemory are marked as \emph{estimated}.  
When a transaction $\mathit{tx}_j$ tries to read an estimated value from a transaction $\mathit{tx}_i$ with $i < j$, the execution is suspended.
Execution of $\mathit{tx}_j$ can resume after the next incarnation of $\mathit{tx}_i$ finishes and replaces the estimated writes with the new ones.
Validation fails if it encounters an estimated write.

\subsection{Compressing deferred counter logs}
\label{sec:system:compression}

\begin{figure}[t]
    \includegraphics[width=\linewidth]{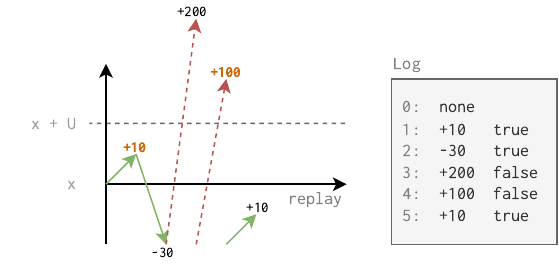}
    \caption{A graphical interpretation of replaying and compressing a log.
    Each entry shifts the starting value \rust{x}.
    Here, the preconditions that the value is always smaller than some limit $U$ (lines 1--2,5 in the log) do not need be re-evaluated.
    Instead, it is sufficient to check $x + 10  \le U$.
    Also, for failed preconditions (lines 3--4) it is enough to check that $x + 80 > U$.}
    \label{fig:deltalane-log-compression}
\end{figure}

With deferred objects, executing transactions produces a set of logs, in addition to a write-set.
Because logs can become large, replaying them at commit time can be slow.
To avoid that, we compress deferred counter logs by taking advantage of the properties of integer addition and the precondition format we support.
Figure~\ref{fig:deltalane-log-compression} shows how logs can be compressed.

It is possible that the first log entry is `\emph{value(v)}'.
In this case, the log contains all the information needed to compress it into a single value by replay.
Similarly, if the log contains a `\emph{revealed}' entry, its value can be used directly.

Otherwise, the log can be compressed into a single summed up update and 4 comparisons.
Consider the $i$-th log entry is an update adding $x_i$.
The precondition for the update is $\lambda v\ .\ L \le v + x_i \le U$.
We let $T$ be all indices of log entries, for which the precondition was true, and $F$ be the indices of entries for which it was false. 
We observe that summing up all updates, i.e., $\sum_{i \in T} x_i$, is equivalent to replaying the log.

Let $x$ be the initial value stored in a deferred counter.
Then adding $x_i$s for $i \in T$ to $x$ just shifts it along the x-axis.
Let $x_{max}$ be the largest shift to the right, and $x_{min}$ the largest shift to the left, and so $[x - x_{min}, x + x_{max}]$ defines the interval where the deferred counter's value lies.
Hence, instead of re-evaluating all preconditions that previously held true, it is sufficient to check that $x + x_{max} \le U$ and $L \le x - x_{min}$.

If we have preconditions that were not satisfied in the log, we need two more comparisons.
A precondition might not be satisfied because adding some $x_i$ for $i \in F$ to the current value $x + \sum_{j \in T, j < i} x_j$ exceeds $U$ or becomes smaller than $L$.
Let $o_{min}$ be the smallest shift to the right such that $U < x + o_{min}$, and $o_{max}$ the largest shift to the left such that $x - o_{max} < L$.
Clearly, it is sufficient to check these two inequalities.

\subsection{Deltas and multi-versioning}
\label{sec:system:deltas}

Instead of using logs directly, \DeltaLane{} uses a more efficient and compact data structure called a \emph{delta}.
Delta is a tuple of
\begin{itemize}
    \item a unique identifier, associated with a deferred object;
    \item a compressed log or a revealed value.
\end{itemize}
A value can be revealed during the log compression if the log contains a \emph{revealed} entry 
  or if the deferred object was newly created (with an initial value).
A delta in a revealed state provides the exact underlying value of the deferred object.
On the other hand, the compressed log contains an \emph{update} (a value to add), and \emph{history constraints} (4 inequalities), and a \emph{source ID} of a deferred object on top of which to apply the updates.
Unless a deferred object is created by a \emph{map} operation, the source ID is equal
  to the ID of the object itself.


A delta with a compressed log can be \emph{applied} to any given base value.
History validation is performed first, i.e., checking that the base value 
  $v$ satisfies the history constraints $h$ of the compressed log.
If this history check fails during delta application, 
  or if during log compression the constraints imposed by \emph{update} and \emph{reveal} operations are not satisfied, 
  the transaction is terminated and the incarnation is marked as speculatively failed.
This can only happen due to speculation in the concurrent setting,
  and means the transaction's internal state is inconsistent with any possible sequential execution.
A speculatively failed incarnation is handled like it had aborted in vanilla Block-STM.

In our implementation, each transaction produces a delta-set as a part of the output, consisting of deltas.
Similar to write-sets, delta-sets are recorded in a dedicated multi-versioned data-structure, 
  \emph{MVDelayedFields}, which is versioned by transaction indices
  and keyed by the deferred object ids. 

\emph{MVDelayedFields} exposes an interface for reading the ``latest'' value of deferred objects.
This requires finding the base value, and potentially applying multiple deltas onto it.
The implementation works by traversing a chain of deltas. 

The \emph{delta traversal} by $\mathit{tx}_i$ of deferred object $A$ starts by finding the 
  highest index $j < i$ with a recorded delta entry $\delta$ in MVDelayedFields for $A$ 
  (analogous for the standard multi-versioned data structure of Block-STM).  
If such a $j$ does not exist, traversal ends with the base value of $A$ from storage.
Similarly, traversal ends if the delta entry of $\mathit{tx}_j$ is contains a revealed \texttt{value}.
Otherwise, the delta has source ID $B$ and the traversal continues by applying 
  the same procedure for $\mathit{tx}_j$ and $B$.
When the base value is found, the deltas are applied to it in the reverse order they were traversed.\footnote{
As an optimization, in cases when $B$ is equal to $A$, 
  the traversal can avoid performing multiple lookups by instead following an ordered iterator.
Additionally, deltas can be merged during the traversal (in a way similar to the log compression), 
  and then applied to the base value.}
  

\begin{figure}[tp]
    \includegraphics[width=\linewidth]{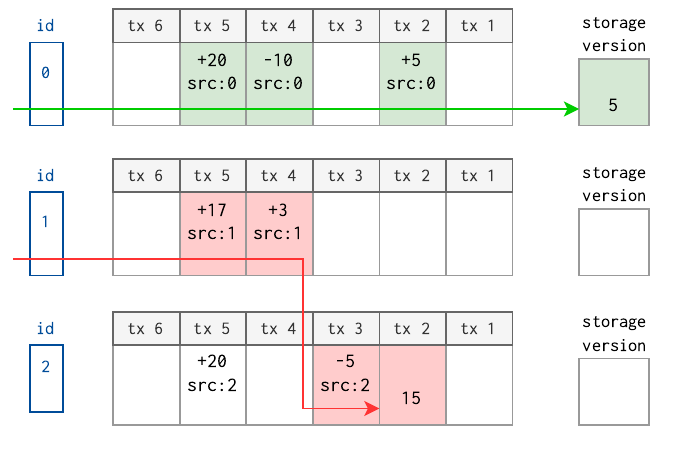}
    \caption{
    Example of the MVDelayedFields data structure for six transactions and three deferred objects.
    Each entry is either an update or a value.
    The arrows represent two delta traversals in MVDelayedFields.
    First, the deferred object with id $0$ is revealed.
    The deltas for the corresponding entry are traversed top-down ($+20$, $-10$, and $+5$) until the storage version is found ($5$), revealing the value of $20$.
    Second, the deferred object with id $1$ is revealed.
    The traversal encounters a redirection to id $2$, and continues the traversal until a value created by transaction $2$ is reached, revealing the value of $30$. }
    \label{fig:system-delta-traversal}
\end{figure}

We add a notable optimization to the Block-STM dependency management 
  for the MVDelayedFields data-structure.
The entries in MVDelayedFields from a particular incarnation are marked as estimates when 
  the incarnation aborts, similar to entries in the multi-versioned data-structure in vanilla Block-STM.
Recall that in Block-STM, validation fails 
upon encountering an estimate.
This would be problematic for \DeltaLane{} as delta traversals visit significantly more entries
in MVDelayedFields.
Hence, we ignore the estimate markings for deferred objects.
Only if a re-execution of a transaction changes the delta entry for a deferred object (or produces a new delta entry), 
the optimization gets disabled and the estimates start being handled as in the vanilla Block-STM for that particular deferred object.

\subsection{Rolling commit}
\label{sec:system:rolling-commit}

Block-STM uses a mechanism called \emph{lazy commit}.
Since the blockchain state is updated per block, lazy commit aims to reduce the synchronization overhead by committing all transaction results in the block at once.
This is a showstopper for an efficient design of \DeltaLane{}, however, as revealing a value, validating a \rust{reveal}, or validating predicted \rust{update} precondition output,
would all require traversing the whole block and accumulating the encountered deltas. 
Another important reason is that the longer the speculative chain is, the less likely the predictions are going to be correct.

To overcome this challenge, we introduce a novel \emph{rolling commit} mechanism.
It allows the system to determine when a transaction will no longer be re-executed, which implies the output is final, and can be committed.

We terminate delta traversals when a committed value is encountered, 
  so the complexity of traversals
  in \DeltaLane{} when executing transaction $\mathit{tx}_i$ depends on $i-j$ as opposed to $i$ 
  when the first $j$ transactions are committed.
This aligns perfectly with the scheduling philosophy that prioritizes tasks with lower $i$ 
(while rolling commit increments $j$).  

Beyond \DeltaLane{}, rolling commit provides fine-grained access to the block execution. 
This is helpful to efficiently support scenarios when only a prefix of transactions needs to be committed, e.g., when the block gas limit\footnote{Maximum allowed amount of gas consumed by all transactions in a block.} is exceeded.
Additionally, block execution is often followed by post-processing tasks based on the committed transaction outputs.
Rolling commit allows these tasks to run in parallel with Block-STM, which can improve the latency of block execution. 

A transaction $\mathit{tx}_i$ can be committed at time $T$ iff:
\begin{itemize}
    \item[(1)] $i=1$ or transaction $\mathit{tx}_{i-1}$ is committed at $T$, and
    \item[(2)] the latest scheduled validation by time $T$ for transaction $\mathit{tx}_i$ has succeeded.
\end{itemize}

As validation can be scheduled for the suffix, it is generally impossible to know if
  a particular validation is the last to ever be scheduled for transaction $\mathit{tx}_i$.
The inductive condition (1) comes into play here,
  as no such validations for $\mathit{tx}_i$ can get scheduled after time $T$, when 
  the prefix $\mathit{tx}_1,\ldots,\mathit{tx}_{i-1}$ is committed.
Checking condition (1) is simple and efficient, e.g., by maintaining an atomic 
  \rust{commit\_idx} counter.
A counter per transaction can be used to check (2), 
  by ordering validations for each transaction based on their scheduling time. 
However, this design requires \rust{schedule\_suffix\_validation(i)} to atomically traverse and update these counters for transactions $\mathit{tx}_i, \ldots, \mathit{tx}_n$, which would be extremely inefficient.
Rolling commit avoids this inefficiency by introducing a global \rust{validation\_wave} counter and instead of updating all validation counters, having validation scheduling check the current validation wave for ordering purposes.
The details of a correct concurrent implementation of such a mechanism are delicate, and are
  explained in~\appendixref{app:rolling-commit}.

\subsection{\DeltaLane{} integration into Block-STM}
In addition to vanilla Block-STM recording all reads during a transaction execution 
  in the CapturedReads component, 
  \DeltaLane{} records all deferred object reveals.
In order to commit an incarnation of $\mathit{tx}_i$ in which a deferred object $A$ was revealed 
  to value $v$, \DeltaLane{} needs to ensure that the value of $A$ is $v$ after the
  first $i-1$ transactions in the block are committed.
Each delta in the delta-set of the incarnation also needs to be validated.
In particular, the history constraint of the delta (in the compressed log)
  needs to be satisfied by the value of the deferred object in the committed state prior to the transaction.
This constraint corresponds to an interval, which is less strict than 
  the equality to a specific value for revealed deferred value,
  and is likely to be satisfied as the output of delta updates are predicted based 
  on the latest committed value of the deferred object. 

\DeltaLane{} performs these deferred object validations right before a transaction is committed by
  the rolling commit mechanism.
The object values can be read efficiently, without traversing any deltas, 
  as the state before the transaction is committed.
Checking the constraints also does not involve any heavy computation.
If deferred object validation succeeds, the transaction gets committed.
If validation fails, 
  the transaction is immediately re-executed and committed without additional validation
  (re-execution happens from the correct state by~\propertyref{prop:bstmvalidation}).
The committed deferred object values get reflected in MVDelayedFields 
  (as deltas with revealed values), 
  allowing the delta traversals to terminate at the committed transaction index.

As discussed previously, for every committed transaction,
Block-STM performs post-commit processing in parallel.
\DeltaLane{} utilizes this processing to incorporate the results of all deferred updates 
into the write-set of the transaction output by inserting inline the committed deferred values. 

An important detail of the above integration is that it requires a slightly stronger theoretical 
  property than the one claimed in the Block-STM paper~\cite{gelashvili2023}.
Validation in vanilla Block-STM compares versions, 
  which are either from storage, or (transaction index, incarnation number)  pairs associated 
  with the execution that produced the value.\footnote{One motivation to compare versions 
    was to avoid performance depending on the (unpredictable) size of values}
Deferred object validation, on the other hand, is based on values.
The subtle difference comes from the fact that incarnation numbers increase monotonically,
  while values could change in the infamous ABA pattern (to a different value that would not pass validation, then back to a value that would).
We verified the proofs in~\cite{gelashvili2023} and ensured that they imply the following:
\begin{property}
Block-STM remains correct if validation compares the read values (instead of the versions) for equality. 
\end{property}

Intuitively, this holds due to~\invariantref{inv:bstmvalidation} and since 
  (a) transaction execution is deterministic given the results of the reads,
  and (b) validation checks all reads, deferred object reveals and delta constraints.

\section{Evaluation}\label{sec:evaluation}

\DeltaLane{} is implemented in Rust, and is integrated into the Block-STM execution engine used by the \Aptos{} blockchain, with the source code being publicly available.
\DeltaLane{} has been released in production as well.

\subsection{Benchmarking methodology}
We used a Google Cloud \emph{t2d-standard-60} (AMD Milan CPU, 60 cores, 240GB memory), hyper-threading disabled.

We evaluated \DeltaLane{} using the benchmarking infrastructure of the \Aptos{} blockchain.
All benchmarks were performed on a single node.
Each benchmark consisted of measuring the throughput of executing and committing 10 blocks, with 10,000 transactions per block.
The global state was initialized with 200,000 accounts in order to mimic a real blockchain setup.
Additionally, 20,000 more accounts were created, each with a sufficiently large balance to pay for transactions.
Each transaction \emph{sender} was selected uniformly at random from these additional accounts.

We consider two sets of workloads: realistic, listed in Table~\ref{tab:evaluation-real-workloads}, and synthetic, listed in Table~\ref{tab:evaluation-artificial-workloads}.
While realistic workloads cover a large variety of real blockchain use cases with a high degree of contention, we also used a small set of hand-crafted contracts in order to stress-test our system.

\begin{table}[t]
    \caption{Realistic transaction workloads.}
    \vspace{-1em}
    \centering
    \small
    \begin{tabular}{
        p{0.2\linewidth}
        p{0.7\linewidth}}
        \toprule
        Name & Description \\
        \midrule
        \emph{no-op} &
        A transaction without any user-specified code. \\
        \emph{sponsored} &
        Same as \emph{no-op}, but instead of the sender paying for the transaction, the fee payer is specified separately.
        Conflicts with transactions with the same fee payer. \\
        \emph{transfer} &
        A peer-to-peer transfer between a pair of accounts.
        Conflicts with transactions with the same sender or the same receiver. \\
        \emph{nft-mint} &
        A transaction that mints an NFT from a collection.
        The collection keeps a counter to track the supply of NFTs minted so far.
        The index of the minted token is used as a part of its name. \\
        \bottomrule
    \end{tabular}
    \label{tab:evaluation-real-workloads}
\end{table}

\begin{table}[t]
    \caption{Synthetic transaction workloads for stress-testing.}
    \vspace{-1em}
    \centering
    \small
    \begin{tabular}{
        p{0.2\linewidth}
        p{0.7\linewidth}}
        \toprule
        Name & Description \\
        \midrule
        \emph{history(n)} &
        Runs a loop to increment a deferred counter $n$ times.
        The increment never overflows. \\
        \emph{cnt(n)} &
        Increments or decrements a deferred counter (chosen at random).
        The bounds of the counter are $[0, n]$.
        The initial value is $0$. \\
        \emph{reveal} &
        Increments and randomly chooses to reveal a deferred counter.
        The counter never overflows. \\
        \bottomrule
    \end{tabular}
    \label{tab:evaluation-artificial-workloads}
\end{table}

\subsection{Realistic workloads}

First, we consider blocks of \emph{no-op} transactions.
Because \Aptos{} tracks the total supply and burns fees for every transaction, even this simple workload is fully sequential.
Figure~\ref{fig:evaluation-real-supply} shows the throughput when supply is tracked using integers, deferred objects, or not tracked at all.
Tracking supply with integers yields a slowdown of up to $6.5\times$ when compared to not tracking it, whereas using deferred objects shows competitive performance.

\pgfplotsset{small,label style={font=\fontsize{1}{2}\selectfont},legend style={font=\fontsize{5}{7}\selectfont},height=3.7cm,width=1.1\textwidth}

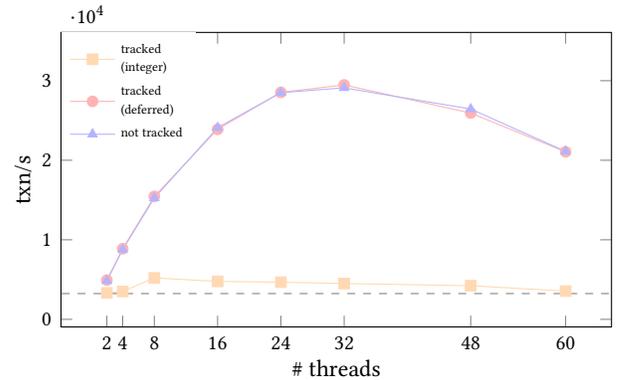
\begin{figure}[tb]
\pgfplotsset{compat=1.18}
    \begin{tikzpicture}
        \begin{axis}[
            legend cell align=left,
            legend columns=1,
            height=5.5cm,
            width=8.9cm,
            ymax=31800,
            ylabel={txn/s},
            xlabel={\# threads},
            enlargelimits=0.25,
            xtick=data,
            enlarge x limits={0.1},
            enlarge y limits={0.15},
            legend style={at={(0,1)}, anchor=north west, draw = none, cells={align=left}},
            xlabel style={font=\small, yshift=3pt},
            ylabel style={font=\small, yshift=-3pt},
        ]

        \addplot [color=orange!30,mark=square*] coordinates {	
            (2, 3317) (4, 3504) (8, 5207) (16, 4772) (24, 4669) (32, 4504) (48, 4228) (60, 3550)
        };
        \addlegendentry{tracked \\ (integer)};
    
        \addplot [color=red!30,mark=*] coordinates {
            (2, 4931) (4, 8878) (8, 15458) (16, 23889) (24, 28538) (32, 29472) (48, 25935) (60, 21061)
        };
        \addlegendentry{tracked \\ (deferred)};

        \addplot [color=blue!30,mark=triangle*] coordinates {
            (2, 4842) (4, 8785) (8, 15253) (16, 24094) (24, 28501) (32, 29104) (48, 26445) (60, 21113)
        };
        \addlegendentry{not tracked};
        
        \addplot [gray, line legend, dashed, sharp plot, update limits=false,shorten >=-6mm,shorten <=-6mm] coordinates {	
            (2, 3240) (4, 3240) (8, 3240) (16, 3240) (24, 3240) (32, 3240) (48, 3240) (60, 3240)
        };

        \end{axis}
    \end{tikzpicture}
    \vspace{-1em}
\caption{Throughput comparison of the \emph{no-op} workload when the total token supply is tracked with an integer, with a deferred counter, or is not tracked at all.
The dashed line shows the throughput of sequential execution.
Deferred counters match the performance of the when the total supply is not tracked at all.
}
\label{fig:evaluation-real-supply}
\end{figure}

In the next experiment, we assume that supply is tracked with a deferred counter, and consider a block of \emph{sponsored} transactions.
Figure~\ref{fig:evaluation-real-fee-payer} shows the throughput for different number of payers in the system, where payers' balances use integer and deferred counters.
With integer balances, the number of read-write conflicts decreases as more payers are used in the system.
With deferred objects, the conflicts are eliminated even for a single payer, and the throughput scales in the same way as in our \emph{no-op} experiment - achieving a speedup of up to $10.3\times$ (compared to integers). 

\pgfplotsset{small,label style={font=\fontsize{1}{2}\selectfont},legend style={font=\fontsize{5}{7}\selectfont},height=3.7cm,width=1.1\textwidth}

\begin{figure}[tb]
\pgfplotsset{compat=1.18}
    \begin{tikzpicture}
        \begin{axis}[
            legend cell align=left,
            legend columns=1,
            height=5.5cm,
            width=8.9cm,
            ylabel={txn/s},
            xlabel={\# threads},
            enlargelimits=0.25,
            xtick=data,
            enlarge x limits={0.1},
            enlarge y limits={0.15},
            legend style={at={(0,1)}, anchor=north west, draw = none, cells={align=left}},
            xlabel style={font=\small, yshift=3pt},
            ylabel style={font=\small, yshift=-3pt},
        ]

        \addplot [color=blue!30,mark=diamond*] coordinates {
            (2, 4653) (4, 8478) (8, 14777) (16, 23467) (24, 27314) (32, 28751) (48, 25528) (60, 19696)
        };
        \addlegendentry{d1};
					
        \addplot [color=yellow!50,mark=square*] coordinates {
            (2, 2582) (4, 2390) (8, 2293) (16, 2359) (24, 2291) (32, 2083) (48, 1903) (60, 1902)
        };
        \addlegendentry{i1};	
 								
        \addplot [color=orange!30,mark=square*] coordinates {
            (2, 4063) (4, 5535) (8, 7083) (16, 7103) (24, 6970) (32, 7008) (48, 6786) (60, 6898)
        };
        \addlegendentry{i4};							
        							
        \addplot [color=purple!30,mark=*] coordinates {
            (2, 4490) (4, 7603) (8, 14778) (16, 14665) (24, 16143) (32, 16973) (48, 17033) (60, 17208)
        };
        \addlegendentry{i16};
	
        \addplot [color=red!30,mark=triangle*] coordinates {
            (2, 4652) (4, 8017) (8, 18241) (16, 18003) (24, 20179) (32, 21598) (48, 21232) (60, 21196)
        };
        \addlegendentry{i32};

        \end{axis}
    \end{tikzpicture}
    \vspace{-1em}
\caption{Throughput comparison for the \emph{sponsored} workload with different number of payers.
In the legend: $iN$ --- $N$ payers with integer balances; $dN$ -- $N$ payers with balances being deferred counters.
Increasing the number of payers when using integer balances improves the throughput.
A single fee payer with balance being a deferred counter outperforms all of them by up to $15.2\times$.}
\label{fig:evaluation-real-fee-payer}
\end{figure}
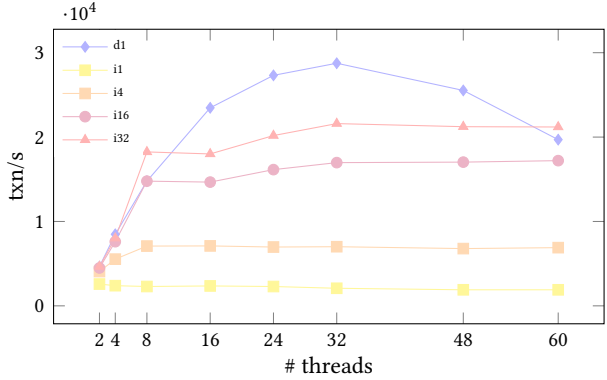

Figure~\ref{fig:evaluation-real-coin-transfer} shows the throughput of peer-to-peer transfers using 1) integer balances and random pairs of accounts, and 2) a single receiver with balances using deferred counters.
Using deferred objects avoids the sequential bottleneck, resulting in speedups of up to $11.6\times$.


Note that using deferred objects does not yield perfect scalability, as one might have expected; however, neither does using non-conflicting transactions.
They both peak when $32$ out of $60$ available cores are used.
This is a limitation of the implementation of the Block-STM execution engine we used, and it is outside of the scope of this work.


\pgfplotsset{small,label style={font=\fontsize{1}{2}\selectfont},legend style={font=\fontsize{5}{7}\selectfont},height=3.7cm,width=1.1\textwidth}

\begin{figure}[tb]
\pgfplotsset{compat=1.18}
    \begin{tikzpicture}
        \begin{axis}[
            legend cell align=left,
            legend columns=1,
            height=5.5cm,
            width=8.9cm,
            ymax = 25000,
            ylabel={txn/s},
            xlabel={\# threads},
            enlargelimits=0.25,
            xtick=data,
            enlarge x limits={0.1},
            enlarge y limits={0.15},
            legend style={at={(0,1)}, anchor=north west, draw = none, cells={align=left}},
            xlabel style={font=\small, yshift=3pt},
            ylabel style={font=\small, yshift=-3pt},
        ]
						
        \addplot [color=orange!30,mark=square*] coordinates {
            (2, 2838) (4, 5338) (8, 8894) (16, 15319) (24, 19398) (32, 19371) (48, 16882) (60, 14455)
        };
        \addlegendentry{random transfer (integer)};							
								

        \addplot [color=blue!30,mark=triangle*] coordinates {
            (2, 3004) (4, 5691) (8, 9976) (16, 16083) (24, 20414) (32, 20418) (48, 18998) (60, 14983)
        };
        \addlegendentry{single receiver (deferred)};
        
        \addplot [gray, line legend, dashed, sharp plot, update limits=false,shorten >=-6mm,shorten <=-6mm] coordinates {	
            (2, 1677) (4, 1677) (8, 1677) (16, 1677) (24, 1677) (32, 1677) (48, 1677) (60, 1677)
        };

        \end{axis}
    \end{tikzpicture}
    \vspace{-1em}
\caption{Throughput comparison of the \emph{transfer} workload for different sender/receiver configurations.
Transfers from a single receiver with a deferred counter balance parallelize perfectly, matching the performance of non-conflicting random transfers.}
\label{fig:evaluation-real-coin-transfer}
\end{figure}
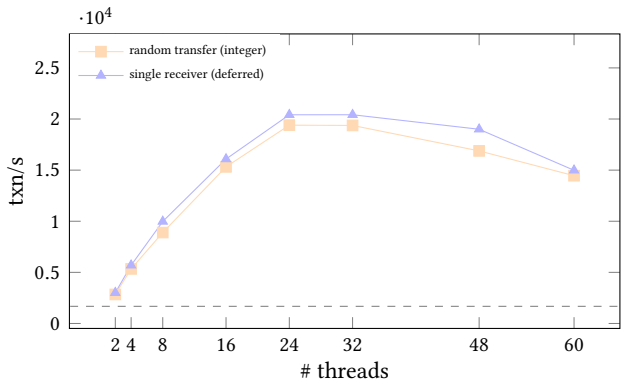

\pgfplotsset{small,label style={font=\fontsize{1}{2}\selectfont},legend style={font=\fontsize{5}{7}\selectfont},height=3.7cm,width=1.1\textwidth}

\begin{figure}[tb]
\pgfplotsset{compat=1.18}
    \begin{tikzpicture}
        \begin{axis}[
            legend cell align=left,
            legend columns=1,
            height=5.5cm,
            width=8.9cm,
            ylabel={txn/s},
            xlabel={\# threads},
            enlargelimits=0.25,
            xtick=data,
            enlarge x limits={0.1},
            enlarge y limits={0.15},
            legend style={at={(0,1)}, anchor=north west, draw = none, cells={align=left}},
            xlabel style={font=\small, yshift=3pt},
            ylabel style={font=\small, yshift=-3pt},
        ]
    
        \addplot [color=red!30,mark=*] coordinates {
            (2, 1866) (4, 3540) (8, 6466) (16, 10702) (24, 13290) (32, 13396) (48, 12083) (60, 9779)
        };
        \addlegendentry{unlimited};		

        \addplot [color=blue!30,mark=triangle*] coordinates {
            (2, 1996) (4, 3654) (8, 3721) (16, 9916) (24, 11787) (32, 9698) (48, 8824) (60, 7814)
        };
        \addlegendentry{limited};
        
        \addplot [gray, line legend, dashed, sharp plot, update limits=false,shorten >=-6mm,shorten <=-6mm] coordinates {	
            (2, 1082) (4, 1082) (8, 1082) (16, 1082) (24, 1082) (32, 1082) (48, 1082) (60, 1082)
        };

        \end{axis}
    \end{tikzpicture}
    \vspace{-1em}
\caption{Throughput comparison of the \emph{nft-mint} workload for unlimited and limited (66,000 NFTs) collections using deferred objects.
The dashed line shows the throughput of the sequential execution when minting an unlimited collection.
With deferred objects the performance is increased in both cases.
Minting the limited collection is only $1.4\times$ worse compared to the unlimited one. 
}
\label{fig:evaluation-nft-mint}
\end{figure}
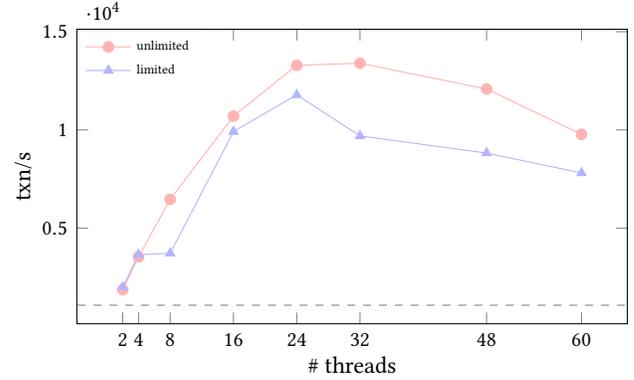

Lastly, we consider an NFT minting workload, with both a limited ($66,000$ NFTs) and unlimited collection.
Both use deferred objects to track the supply and represent the NFT names.
Figure~\ref{fig:evaluation-nft-mint} shows the obtained throughput.
When the collection is limited, \DeltaLane{} estimates if minting succeeds or not.
The estimation is correct nearly at all times, as demonstrated by the high throughput for the limited collection minting, with a speedup of up to $10.1\times$ compared to the sequential baseline.

\subsection{Synthetic workloads}

We first study how mispredicting preconditions affects the throughput, using a block of \emph{cnt} transactions, each randomly incrementing or decrementing a deferred counter.
We vary the upper bound between $1$ and $1000$, and show the throughput in Figure~\ref{fig:evaluation-synthethic-misprediction}.

\pgfplotsset{small,label style={font=\fontsize{1}{2}\selectfont},legend style={font=\fontsize{5}{7}\selectfont},height=3.7cm,width=1.1\textwidth}

\begin{figure}[tb]
\pgfplotsset{compat=1.18}
    \begin{tikzpicture}
        \begin{axis}[
            legend cell align=left,
            height=5.5cm,
            width=8.9cm,
            ylabel={txn/s},
            xlabel={\# threads},
            enlargelimits=0.25,
            xtick=data,
            enlarge x limits={0.1},
            enlarge y limits={0.15},
            legend style={at={(0,1)}, anchor=north west, draw = none, cells={align=left}},
            xlabel style={font=\small, yshift=3pt},
            ylabel style={font=\small, yshift=-3pt},
        ]
        
        \addplot [color=orange!30,mark=square*] coordinates {	
            (2, 3823) (4, 4962) (8, 5183) (16, 4540) (24, 4392) (32, 4526) (48, 4223) (60, 4094)
        };
        \addlegendentry{cnt(1)};
    
        \addplot [color=red!30,mark=*] coordinates {
            (2, 4568) (4, 7678) (8, 8902) (16, 8107) (24, 6824) (32, 7829) (48, 7131) (60, 6662)
        };
        \addlegendentry{cnt(10)};

        \addplot [color=purple!30,mark=triangle*] coordinates {
            (2, 4581) (4, 8396) (8, 13798) (16, 17211) (24, 17565) (32, 16246) (48, 15233) (60, 12592)
        };
        \addlegendentry{cnt(100)};

        \addplot [color=blue!30,mark=diamond*] coordinates {
            (2, 4638) (4, 8463) (8, 14520) (16, 22475) (24, 25603) (32, 26091) (48, 21101) (60, 18863)
        };
        \addlegendentry{cnt(1,000)};

        \addplot [gray, line legend, dashed, sharp plot, update limits=false,shorten >=-6mm,shorten <=-6mm] coordinates {	
            (2, 2950) (4, 2950) (8, 2950) (16, 2950) (24, 2950) (32, 2950) (48, 2950) (60, 2950)
        };
    
        \end{axis}
    \end{tikzpicture}
    \vspace{-1em}
\caption{Throughput comparison of the \emph{cnt(n)} workload, varying the bounds for the deferred counter.
The dashed line shows the throughput of the sequential execution for \emph{cnt(1)}.
When the value of the counter is mispredicted the most ($n = 1$), the throughput is still higher than that of the sequential baseline ($1.8\times$).}
\label{fig:evaluation-synthethic-misprediction}
\end{figure}
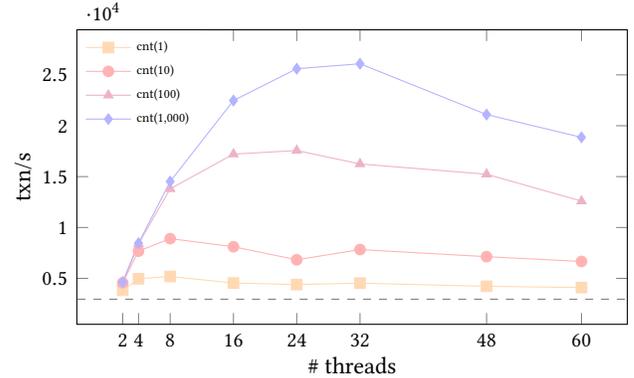

As expected, as the misprediction probability increases, the throughput decreases.
Even in the worst case (upper bound of $1$), with only 50\% chance of predicting correctly we obtain a speedup of up to $1.8\times$.

\pgfplotsset{small,label style={font=\fontsize{1}{2}\selectfont},legend style={font=\fontsize{5}{7}\selectfont},height=3.7cm,width=1.1\textwidth}

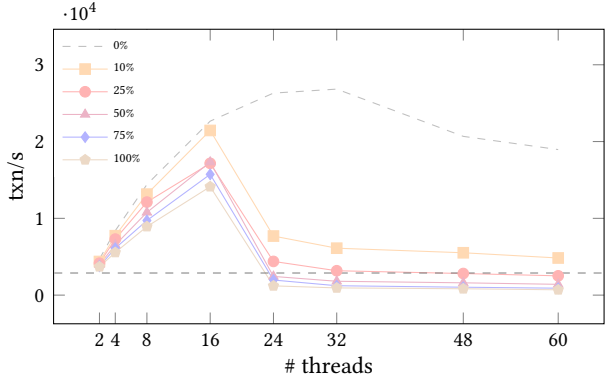
\begin{figure}[tb]
\pgfplotsset{compat=1.18}
    \begin{tikzpicture}
        \begin{axis}[
            legend cell align=left,
            height=5.5cm,
            width=8.9cm,
            ymax=30200,
            ylabel={txn/s},
            xlabel={\# threads},
            enlargelimits=0.25,
            xtick=data,
            enlarge x limits={0.1},
            enlarge y limits={0.15},
            legend style={at={(0,1)}, anchor=north west, draw = none},
            xlabel style={font=\small, yshift=3pt},
            ylabel style={font=\small, yshift=-3pt},
        ]	
							
        \addplot [color=gray!50, dashed] coordinates {	
            (2, 4705) (4, 8391) (8, 14403) (16, 22658) (24, 26313) (32, 26829) (48, 20676) (60, 18956)
        };
        \addlegendentry{0\%};							

        \addplot [color=orange!30,mark=square*] coordinates {	
            (2, 4379) (4, 7741) (8, 13150) (16, 21446) (24, 7692) (32, 6111) (48, 5516) (60, 4836)
        };
        \addlegendentry{10\%};
    							
        \addplot [color=red!30,mark=*] coordinates {
            (2, 4157) (4, 7297) (8, 12116) (16, 17162) (24, 4389) (32, 3175) (48, 2813) (60, 2512)
        };
        \addlegendentry{25\%};

        \addplot [color=purple!30,mark=triangle*] coordinates {
            (2, 3969) (4, 6531) (8, 10777) (16, 17320) (24, 2434) (32, 1804) (48, 1601) (60, 1405)
        };
        \addlegendentry{50\%};

        \addplot [color=blue!30,mark=diamond*] coordinates {
            (2, 3751) (4, 6104) (8, 9719) (16, 15732) (24, 1966) (32, 1223) (48, 1029) (60, 901)
        };
        \addlegendentry{75\%};
				
        \addplot [color=brown!30,mark=pentagon*] coordinates {
            (2, 3693) (4, 5555) (8, 8925) (16, 14122) (24, 1205) (32, 939) (48, 828) (60, 693)
        };
        \addlegendentry{100\%};

        \addplot [gray, line legend, dashed, sharp plot, update limits=false,shorten >=-6mm,shorten <=-6mm] coordinates {	
            (2, 2879) (4, 2879) (8, 2879) (16, 2879) (24, 2879) (32, 2879) (48, 2879) (60, 2879)
        };
    
        \end{axis}
    \end{tikzpicture}
    \vspace{-1em}
\caption{Throughput comparison for the \emph{materialize} workload.
Dashed lines show parallel (top) and sequential (bottom) baselines for $0\%$ materializing transactions.
Using deferred objects introduces an overhead for materialize-intensive workloads for a large number of threads.
}
\label{fig:evaluation-synthethic-reveal}
\end{figure}

Next, we use \emph{reveal} transactions.
Transactions in the block are generated in such a way that a fixed fraction of them reveals a deferred object.
Figure~\ref{fig:evaluation-synthethic-reveal} shows the throughput.
As expected, we observe that revealing the deferred object leads to lower performance.
For instance, when 10\% of transactions do a reveal, the throughput is up to $4.4\times$ lower than when not revealing at all.
As the percentage of revealing transactions increases, \DeltaLane{} performs more traversals, bringing the throughput closer, and sometimes even worse (by at most $3.1\times$), than the sequential baseline.
At the same time, because delta traversals are only performed until the latest committed transaction, there are smaller regressions when fewer threads are used.

In our next experiment, we demonstrate how log compression improves performance when applying deferred computations at commit time.
We use the \emph{history(n)} transaction for that, which would generate a log with $n$ entries without compression.
We plot the increase in throughput (as compared to the baseline of $1$) in Figure~\ref{fig:evaluation-synthethic-compression}.

We observe that the size of the log does not influence the performance, as large logs performed identical or better than small ones.
Compressing logs using deltas is an effective way to avoid excessive computations at commit time.

\pgfplotsset{small,label style={font=\fontsize{1}{2}\selectfont},legend style={font=\fontsize{5}{7}\selectfont},height=3.7cm,width=1.1\textwidth}

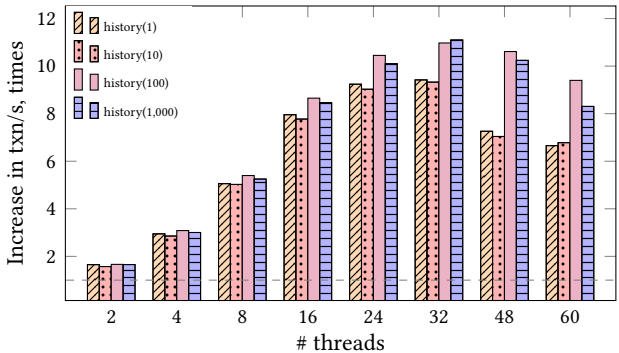
\begin{figure}[tb]
\pgfplotsset{compat=1.18}
    \begin{tikzpicture}
        \begin{axis}[
            ybar,
            bar width=4.5pt,
            /pgfplots/ybar=0pt,
            legend cell align=left,
            height=5.5cm,
            width=8.9cm,
            ylabel={Increase in txn/s, times},
            xlabel={\# threads},
            enlargelimits=0.25,
            symbolic x coords={2, 4, 8, 16, 24, 32, 48, 60},
            xtick=data,
            xtick align=inside,
            enlarge x limits={0.1},
            enlarge y limits={0.15},
            legend style={at={(0,1)}, anchor=north west, draw = none},
            xlabel style={font=\small, yshift=3pt},
            ylabel style={font=\small, yshift=-3pt},
        ]
        
        \addplot [fill=orange!30, postaction = {pattern = north east lines}] coordinates {				
            (2, 1.652036517) (4, 2.94627809) (8, 5.057233146) (16, 7.955758427) (24, 9.239115169) (32, 9.420294944) (48, 7.259831461) (60, 6.655898876)
        };
        \addlegendentry{history(1)};
    
        \addplot [fill=red!30, postaction = {pattern = dots}] coordinates {			
            (2, 1.572347267) (4, 2.855305466) (8, 5.024294391) (16, 7.773133262) (24, 9.027152554) (32, 9.326902465) (48, 7.041086102) (60, 6.781350482)
        };
        \addlegendentry{history(10)};

        \addplot [fill=purple!30] coordinates {			
            (2, 1.663456665) (4, 3.084135834) (8, 5.397871262) (16, 8.652812975) (24, 10.45007603) (32, 10.97263051) (48, 10.61226559) (60, 9.400405474)
        };
        \addlegendentry{history(100)};

        \addplot [fill=blue!30, postaction = {pattern = horizontal lines}] coordinates {			
            (2, 1.655308642) (4, 3.00345679) (8, 5.250864198) (16, 8.460740741) (24, 10.10123457) (32, 11.09876543) (48, 10.24098765) (60, 8.304691358)
        };
        \addlegendentry{history(1,000)};

        \addplot[gray, line legend, dashed, sharp plot, update limits=false,shorten >=-6mm,shorten <=-6mm] coordinates {
            (2, 1) (4, 1) (8, 1) (16, 1) (24, 1) (32, 1) (48, 1) (60, 1)
        };
    
        \end{axis}
    \end{tikzpicture}
\vspace{-1em}
\caption{Throughput improvement for the \emph{history(n)} workload when compressing logs (higher is better).
}
\label{fig:evaluation-synthethic-compression}
\end{figure}

\subsection{Rolling commits}
\label{sec:eval:rolling-commit}

Finally, we evaluate another important contribution of this paper: rolling commits. 
We compare it with lazy commit, as used by the original Block-STM.
We use the same benchmark and setup as in the original Block-STM paper~\cite{gelashvili2023}.
Figure~\ref{fig:eval:rolling-commit} shows the throughput of executing and committing \emph{transfer} transactions using the two commit mechanisms.

\pgfplotsset{small,label style={font=\fontsize{1}{2}\selectfont},legend style={font=\fontsize{5}{7}\selectfont},height=3.7cm,width=1.1\textwidth}

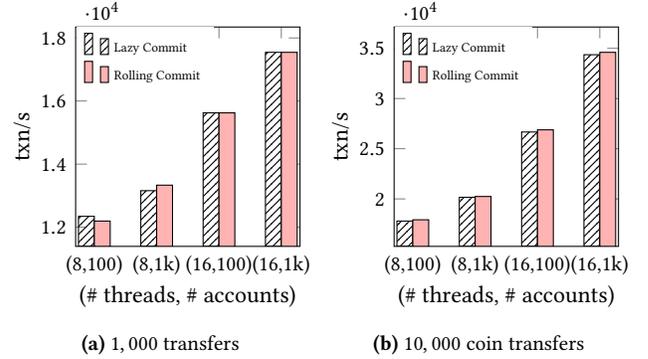
\begin{figure}[tb]
\pgfplotsset{compat=1.18}
\begin{minipage}{\columnwidth}
    \begin{subfigure}{0.49\columnwidth}
        \begin{tikzpicture}
            \begin{axis}[
                legend cell align=left,
                ybar,
                height=4.5cm,
                ylabel={txn/s},
                xlabel={(\# threads, \# accounts)},
                enlargelimits=0.25,
                bar width=6pt,
                /pgfplots/ybar=0pt,
                symbolic x coords={{(8,100)}, {(8,1k)}, {(16,100)}, {(16,1k)}},
                xtick=data,
                xtick align=inside,
                enlarge x limits={0.1},
                enlarge y limits={0.15},
                legend style={at={(0,1)}, anchor=north west, draw = none},
                xlabel style={font=\small, yshift=3pt},
                ylabel style={font=\small, yshift=-3pt},
            ]
            \addplot [pattern = north east lines] coordinates {({(8,100)}, 12345) ({(8,1k)}, 13158) ({(16,100)}, 15625) ({(16,1k)}, 17544)};
            \addlegendentry{Lazy Commit};
        
            \addplot [fill=red!30!white] coordinates {({(8,100)}, 12195) ({(8,1k)}, 13333) ({(16,100)}, 15625) ({(16,1k)}, 17544)};
            \addlegendentry{Rolling Commit};
        
            \end{axis}
        \end{tikzpicture}
        \subcaption{$1,000$ transfers}
        \label{fig:eval:rc:1}
    \end{subfigure}
    \begin{subfigure}{0.49\columnwidth}
        \begin{tikzpicture}
            \begin{axis}[
                legend cell align=left,
                ybar,
                height=4.5cm,
                ylabel={txn/s},
                xlabel={(\# threads, \# accounts)},
                enlargelimits=0.25,
                bar width=6pt,
                /pgfplots/ybar=0pt,
                symbolic x coords={{(8,100)}, {(8,1k)}, {(16,100)}, {(16,1k)}},
                xtick=data,
                xtick align=inside,
                enlarge x limits={0.1},
                enlarge y limits={0.15},
                legend style={at={(0,1)}, anchor=north west, draw = none},
                xlabel style={font=\small, yshift=3pt},
                ylabel style={font=\small, yshift=-3pt},
            ]
            \addplot [pattern = north east lines] coordinates {({(8,100)}, 17794) ({(8,1k)}, 20161) ({(16,100)}, 26667) ({(16,1k)}, 34364)};
            \addlegendentry{Lazy Commit};
        
            \addplot [fill=red!30!white] coordinates {({(8,100)}, 17921) ({(8,1k)}, 20242) ({(16,100)}, 26881) ({(16,1k)}, 34602)};
            \addlegendentry{Rolling Commit};
        
            \end{axis}
        \end{tikzpicture}
        \subcaption{$10,000$ coin transfers}
        \label{fig:eval:rc:2}
    \end{subfigure}
    \vspace{-1em}
    \caption{Throughput comparison of \emph{transfer} with different commit algorithms.
    Contention is determined by the number of accounts in the system (smaller is higher).
}
    \label{fig:eval:rolling-commit}
\end{minipage}
\end{figure}

Lazy and rolling commits achieve similar performance in all cases, proving that the synchronization overhead introduced by rolling commit is minimal.
Hence, rolling commit provides significantly expanded functionality requiring additional fine-grained synchronization at a negligible performance cost.
Perhaps surprisingly, rolling commit has better performance in some cases, which is explained by the post-processing improvements, since lazy commit has to wait until the whole block finishes execution before post-processing, which offsets rolling commit overhead.

\section{Related Work}\label{sec:related-work}

There has been much prior work on executing smart contracts in parallel using both pessimistic~\cite{wei2018, bartoletti2020, lu2023} and optimistic~\cite{anjana2019, gelashvili2023, qi2023} approaches, but less attention has been given to optimizing the execution of workloads with inherent read-write conflicts~\cite{saraph2019, garamvolgyi2022}.

Some existing techniques exploit commutativity of operations~\cite{pirlea2021, qi2023, miller2024} -- a well-known technique when implementing efficient concurrent data structures~\cite{herlihy1995, kulkarni2011}.
Sharding counters~\cite{herlihy1995, lloyd2016} is another technique used to ensure collision-free modifications of shared data.
However, these techniques are not directly applicable in blockchains where execution has to be deterministic.

There exist many language abstractions for writing concurrent programs~\cite{coarsepopl, galois, futures, transaction-boosting, scheme-futures,dean2008,tvm}.
Loop reductions updating a shared variable also have been extensively studied~\cite{openmp, cilk-reducers},
including automatic loop vectorization done by compilers~\cite{siso19,poly10}.
There is also work on creating concurrency-aware types~\cite{burckhardt2010, holt2016}.

The concept of deferring computations has also been used in other domains, including machine learning frameworks~\cite{suhan2021, lopes2023}, concurrent data structure design~\cite{afek2013, moir2018}, and databases~\cite{gawlick1985, rocksdbwiki2024}.
However, we are unaware of any language concept similar to deferred objects used to reduce concurrency conflicts. 

\section{Conclusion}\label{sec:conclusion}

We presented deferred objects, a new construct for smart contract programming languages that enables conflicting transactions (with read-write conflicts) to be run in parallel.

We designed and implemented an extension to the Block-STM parallel execution engine to support deferred objects, called \DeltaLane{}.
We integrated \DeltaLane{} in the \Aptos{} blockchain and showed that it can improve the throughput of real workloads by up to $12\times$.

\bibliographystyle{ACM-Reference-Format}
\bibliography{references}
\appendix

\section{Rolling Commit Details}

\label{app:rolling-commit}

Each rolling commit has a global \rust{validation\_wave} counter, which is incremented by \rust{schedule\_suffix\_validation(i)}, i.e., it triggers a new validation wave.
When validation starts, it reads the \rust{validation\_wave} counter value $w$.
When transaction $\mathit{tx}_i$ gets scheduled for validation,
  the value $w$ is recorded for $i$ 
  as $\rust{required\_wave[i]} \gets \max(\rust{required\_wave[i]}, w)$.\footnote{Maximum is not 
  needed if the read of \rust{validation\_wave} and update of \rust{required\_wave} 
  happens atomically, but \rust{validation\_wave} is kept separate, as a more contended counter, to minimize synchronization overhead.} 

Instead of checking condition (2), rolling commit considers the maximum wave $w$ for which 
  validation of transaction $\mathit{tx}_i$ has succeeded, and checks whether $w$ is greater or equal to 
  both \rust{required\_wave[i]} and the maximum validation wave triggered at indices $1, \ldots, i$.
Recall that by design in Block-STM, 
  there can be multiple concurrent validations of the same transaction.
\rust{validation\_wave} is used as a timer, and the above check 
  can diverge from correctness condition (2) outlined in~\sectionref{sec:system:rolling-commit} to consider success of a validation $v'$ 
  different from the validation $v$ that was scheduled last, iff $v'$ started later than $v$.
This is correct as $v'$ will also validate from the global state corresponding to 
  the sequential execution of prior transactions (\propertyref{prop:bstmvalidation}).
Finally, each time a new wave is triggered by \rust{schedule\_suffix\_validation(i)},
  it is recorded in $\rust{triggered\_wave[i]}$ (similar to required wave).
This allows rolling commit to trivially maintain the sweeping maximum triggered 
  wave number for the committed prefix.

\begin{figure}[t]
    \includegraphics[width=\linewidth]{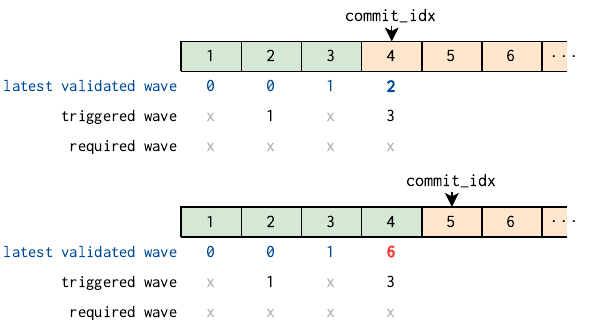}
    \caption{Example of rolling commit trying to commit transaction $4$. 
In the first figure, rolling commit will not succeed since the latest validated wave
  is $2$, which is less than the maximum triggered wave for transaction $4$ 
  (which is equal to $3$ and computed as the maximum of triggered waves for transactions $1,2,3$ and $4$).
At a later point, when current validation wave is $6$, transaction $4$ gets re-validated 
  again and the latest validated wave gets updated to $6$, allowing transaction $4$ 
  to be committed.
  }
    \label{fig:deltalane-rolling-commit}
\end{figure}

A simple example of rolling commit (without required wave being set) is given in~\figureref{fig:deltalane-rolling-commit}.
A more detailed specification of the protocol is as follows:

\begin{itemize}
    \item \cidx is an atomic counter that records the index of the next transaction to be committed, which gets incremented each time when a transaction gets committed.
    \item \vwave is an atomic counter that records the current validation wave, which gets incremented each time \rust{schedule\_suffix\_validation} is called.
    \item For transaction $i$, the system tracks three atomic counters \trwave, \ttwave and (latest) \tvwave, which gets updated as follows:
    \begin{itemize}
        \item $\trwave\gets \max(\trwave, \\ \vwave)\\$ when \rust{schedule\_validation(i)} is called, meaning that transaction $i$ needs 
        to be successfully validated by a wave $\geq \trwave$.
        \item $\ttwave\gets \max(\ttwave, \\ \vwave)$ when \\ \rust{schedule\_suffix\_validation(i)} is called, meaning that any transaction $\geq i$ will need to be validated by a wave $\geq \ttwave$.
        \item $\tvwave\gets \max(\tvwave, w)$ \\ when the a validation succeeds, where $w$ is the wave of the 
        validation, assigned at start by reading the \vwave counter.
    \end{itemize}
    \item \cwave is an atomic counter that records the highest required wave of a successful validation for the next transaction to be committed, updated as \\ $\cwave\gets\max(\cwave, \\ \ttwave[i])$ 
    when trying to commit a transaction $i$.
\end{itemize}

The scheduler updates \vwave and \cwave properly during re-validation and commit. To commit the next transaction, the scheduler needs to ensure the validation task is successful, and that the validation started sufficiently late, i.e. belonged to a sufficiently large wave.
This is accomplished by checking \tvwave against \cwave and \trwave.

\balance

More specifically, 
    the transaction \cidx is committed $\tvwave\geq \max(\cwave, \trwave)$ holds for it, and \cidx is incremented by 1.

Compared to lazy commit, the only additional overhead may come from the cost of maintaining the wave information. To minimize the overhead, our implementation concatenates the wave number with the validation index to be the new validation index, and use atomic operations to update the wave number or validation index.
Our evaluation in Section~\ref{sec:eval:rolling-commit} demonstrates that rolling commit adds no overhead compared to lazy commit.

An alternative approach would be to dedicate one thread to incrementally validate and commit transactions.
Note that this adds one extra validation (one that succeeds in order to commit) per transaction.
More importantly, when Block-STM is used with few threads, losing one thread is significant.
However, when Block-STM is used with many threads, it is possible for the committing thread to 
  become the latency bottleneck.
In contrast, rolling commit interprets the results of validations organically scheduled by vanilla Block-STM
  with careful synchronization to determine when transactions can be committed.

\end{document}